%% file: thalf_fp.tex
\documentclass[epj]{svjour}
\usepackage{graphicx,epstopdf,multirow,amsmath,rotating,subfigure,isotope,gensymb,amssymb}
\usepackage{graphicx}
\usepackage{dcolumn}
\usepackage{bm}
\usepackage{tabularx}
\usepackage[export]{adjustbox}
\usepackage{float}
\usepackage{braket}
\usepackage{lscape}
\usepackage{lipsum}
\usepackage{longtable}

\usepackage{latexsym}
\usepackage{graphics}
\usepackage{xcolor}   

\usepackage{hyperref}
\hypersetup{
    colorlinks=true, 
    linktoc=all,     
    urlcolor  = blue,
    citecolor = blue,
    linkcolor=blue,  
}
\makeatletter
\newcommand{\colorcaption}[2][]{%
  \begingroup%
  \renewcommand{\@caption@fignum@sep}{ (Color online). }%
  \caption[#1]{#2}%
  \endgroup%
}
\makeatother
\usepackage{amsmath}
\begin{document}
\title{Shell-model study of  $\beta^+$/EC-decay half-lives for $Z = 21-30$ nuclei}
\author{Vikas Kumar$^1$\thanks{vikasphysics@bhu.ac.in}, Praveen C.~Srivastava$^2$\thanks{praveen.srivastava@ph.iitr.ac.in}}
\institute{$^{1}$Department of Physics, Institute of Science, Banaras Hindu University Varanasi, Varanasi - 221 005, INDIA \\
$^{2}$Department of Physics, Indian Institute of Technology Roorkee, Roorkee 247 667, 
INDIA }

\date{\today}

\abstract{In the present work, we have reported  $\beta^+$/EC-decay half-lives for $Z = 21 -30$ nuclei using large-scale shell-model. 
A recent study shows that some proton-rich nuclei in this region belong to the island of inversion. We have performed calculations for these nuclei using KB3G effective interaction, while for Ni, Cu, and Zn nuclei we have used JUN45 effective interaction in the  $f_{5/2}pg_{9/2}$ model space.
The calculated quenching factors for $fp$ and $f_{5/2}pg_{9/2}$ space using KB3G and JUN45 effective interactions are also reported. Shell-model results of $\beta$-decay half-lives, excitation energies, log$ft$ values, and branching fractions are discussed and compared with the available experimental data. We have obtained a reasonable agreement with the available data.}

\authorrunning{Vikas Kumar and P.C. Srivastava}
\maketitle


\section{Introduction}
The recent development of radioactive ion facilities opens opportunities to study nuclei away from the
stability lines, these nuclei may decay by different modes of beta decay. Thus, it is important to study
these nuclei theoretically using novel approaches. Recently, beta decay properties using ab initio theory
were reported in Ref. \cite{nature}. It is still very challenging to study nuclei in the entire region of the
nuclear chart using an ab initio approach, thus study of these nuclei using a large-scale shell model is
very important. A weak sub-shell effect at $N$=40 in the Cu isotopes is obtained in \cite{M. L. Bissell}. Recchia {\it et al.} \cite{F Recchia} reported sub-shell closure in Co isotopes toward the $N$=40 and the new island of inversion. The recent studies in  \cite{vikas,prc8161301,vikas1,prc8564305,vikas2,prc8614325,pc,prl102142501,prl115192501} show the coexistence of normal and intruder configurations in neutron rich nuclei around $N$=40 shell gap. Due to the emergence of  a new island of inversion and sub-shell closure around $N$=40 the nuclear structure study including $\beta$-decay properties of these nuclei is very important \cite{Otsuka,p3104,williams,prc8254301}. 
Recently, shell-model results for neutron capture rates for several nuclei including Cr and Ni chains are reported in Ref. \cite{Sieja}.

The experimental $\beta$-decay half-lives of $^{41,42}$Sc nuclei are reported in \cite{41sc,42sc}. The spectrum of $\gamma$-rays following the $\beta$-decay of $^{43}$Sc has been measured and reported in \cite{43sc}.  Sarriguren {\it et al.} \cite{Sarriguren} studied stellar electron-capture rates  for $fp$ shell nuclei using quasiparticle random-phase approximation. 
In \cite{scvmnco}, the half-lives of $^{42}$Sc, $^{46}$V, $^{50}$Mn, and $^{54}$Co nuclei are reported.
The ground-state $\beta^+$-decay of $^{42}$Ti has been investigated and the half-life, branching ratios, and log$ft$ values are reported in \cite{42ti}. The beta decay of $^{43}$Ti to its mirror nucleus $^{43}$Sc has been studied up to 5 MeV excitation energy in \cite{43ti}. First time, observations of non-analog $0^+$$\rightarrow$$0^+$ branches in $^{38m}$K, $^{46}$V, $^{50}$Mn, and $^{54}$Co are reported in \cite{46v}. 
The $\beta$-delayed proton radioactivity of proton-rich nuclei $^{44}$Cr, $^{47}$Mn, $^{48,49}$Fe and $^{50}$Co are studied in \cite{44v}. In Dossat {\it et al} \cite{44v1}, a series of experiments at the SISSI/LISE3 facility of GANIL was conducted for long time and they reported $\beta$-decay half-life and their total $\beta$-delayed proton emission branching ratio of 26 nuclei between $^{36}$Ca and $^{56}$Zn. Orrigo {\it et al} \cite{48fe} reported the half-lives and the total $\beta$-delayed proton emission branching ratios of three proton-rich nuclei with $T_{z}$ = -2, namely $^{48}$Fe, $^{52}$Ni, and $^{56}$Zn, produced at GANIL. The $\beta^+$-decay study of $T_z$ =-1 nuclei $^{54}$Ni, $^{50}$Fe, $^{46}$Cr, and $^{42}$Ti produced in fragmentation reactions at GSI is reported by Molina {\it et al} \cite{50fe}. The first time the $\beta$-decay half-life of $^{58}$Zn has been determined at the ISOLDE on-line separator facility at CERN in \cite{58zn}. In \cite{64cu}, half-lives and $\gamma$-ray intensities of $^{64}$Cu and $^{68}$Ga were measured at IFIN-HH.

In the last few years, many experiments have been performed around the globe using RIB facilities for the measurement of $\beta$-decay half-lives, log$ft$ values, branching ratios, etc. in $fp$ and $f_{5/2}pg_{9/2}$ shell nuclei, many of them are neutron rich nuclei and belongs to the island of inversion. A systematic theoretical estimate for $\beta^+$-decay half-lives of neutron rich nuclei is needed despite the progress in the experimental side.
Motivated with these recent data, in the present work we have performed a systematic shell-model (SM) study of 
$\beta$-decay half-lives, excitation energies, log$ft$ values, and branching fractions for $Z = 21 - 30$ nuclei.
In the present work, our SM calculations are based on allowed  Fermi and  GT-transitions.
Previously, SM calculations for the $\beta^-$-decays of $fp$ and $fpg$ shell nuclei are reported by us in Ref.  \cite{vikas}.

This paper is organized as follows. In section 2, the theoretical formulas for $\beta$ decay half-lives calculations are discussed. Shell model spaces, effective interactions and the quenching factor calculations
 are reported in section 3. In section 4, the theoretical results along with the experimental data are discussed. Finally, in section 5 the summary and conclusions are drawn.

\section{Formalism for $\beta$-decay half-lives}
In all three different types of $\beta$-decay processes, the mass number A of the parent nucleus remains unchanged, only the atomic number Z changes by one unit. The $\beta$-decay selection rules permit all those transitions which are inside the energy window defined by $Q$-value from the ground state of the parent nucleus to different excited states of the daughter nucleus. 
The transition probability $T_{fi}$ is related to the half-life of $\beta$-decay as

\begin{equation}
 t_{1/2}=\frac{ln2}{T_{fi}}.
\end{equation}

The resulting expression for total decay half-life of a combined $\beta^+$ and electron-capture (EC) transition, denoted by $\beta^+$/EC, is given by
\begin{equation}
 f_{0}t_{1/2}=\left[f_0^{(+)}+f_0^{(EC)}\right]t_{1/2} = \frac{\kappa}{[g^2_A*B(GT) + B(F)]},
\end{equation}
where, $g_A$ (= 1.270) represents the axial-vector coupling constant of the weak interactions and $f_0$ represents the phase-space factor, sometimes also called the Fermi integral. The B(F) and B(GT) are the Fermi and Gamow–Teller reduced transition probabilities, respectively. the latest updated value of $\kappa$ is taken from \cite{Patrignani}

\begin{equation}
 \kappa \equiv \frac{2\pi^3\hbar^7ln2}{m^5_ec^4{(G_Fcos{\theta}_C)}^2} = 6289s,
\end{equation}
where, the ${\theta}_C$ is the Cabibbo angle. 

The Fermi reduced transition probability $B(F)$ is given by
\begin{equation}
 B(F) \equiv \frac{g^2_V}{{2J_i + 1}}{|M_F|^2},
\end{equation}
where, $g_V$ (= 1.0) represents the vector coupling constant of the weak interaction and $M_F$ is the Fermi matrix element.

The Gamow-Teller reduced transition probability $B(GT)$ is given by
\begin{equation}
 B(GT)= {\langle{\sigma\tau}\rangle}^2.
\end{equation}
 In the above expression, the nuclear matrix element for the Gamow-Teller operator is given by
\begin{equation}
 {\langle{\sigma\tau}\rangle} = \frac{{\langle {f}|| \sum_{k}{\sigma^k\tau_{\pm}^k} ||i \rangle}}{\sqrt{2J_i + 1}},
 \label{sigmatau}
\end{equation}
where initial and final states are represented by the quantum numbers $i$ and $f$, respectively.
$\pm$ refers to $\beta^{\pm}$ decay, $\tau_{\pm} = \frac{1}{2}(\tau_x + i\tau_y)$ with $\tau_+p$ = $n$, $\tau_-n$ = $p$, and $J_i$ is the total angular momentum of the initial-state. The sum in Eq. \ref{sigmatau} runs over all the nucleons.

 For $\beta^{\mp}$ decay, the phase-space factor is given by
 
\begin{equation}
 f_0^{(\mp)}= \int_{1}^{E_0} F_0({\pm}Z_f, \epsilon)p\epsilon(E_0-\epsilon)^2 \,d\epsilon,
\end{equation}
where, $F_0 $ is called Fermi function and 
\begin{equation}
 \epsilon \equiv \frac{E_e}{m_ec^2}, E_0 \equiv \frac{E_i-E_f}{m_ec^2}, p \equiv \sqrt{\epsilon^2-1},
\end{equation}
where $E_e$ is the total energy of the emitted electron/positron and $E_i$ and $E_f$ are the energies of the initial and final nuclear state.\\
    The phase-space factor for the electron capture \cite{suhonen} is given by
\begin{equation}
f^{(EC)}_0 = 2\pi({\alpha}Z_i)^3(\epsilon_0+E_0)^2,
\end{equation}
where 
\begin{equation}
\epsilon_0 \equiv 1-\frac{1}{2}({\alpha}Z_i)^2,
\end{equation}
and $\alpha$ is the fine-structure constant, ${\alpha}=\frac{1}{137}$. The simple non-relativistic $s$-electron wave function was assumed in Eq. 10. This approximation is valid when $\alpha Z_i$ $\ll$1. For $Z_i$ $<$ 40 this approximation holds good.
In a non-relativistic approximation the Fermi function $F_0$ can be written as Primakoff-Rosen approximation \cite{Primakoff}
\begin{equation}
F_0(Z_f, \epsilon) \approx \frac{\epsilon}{p}F^{(PR)}_0(Z_f).
\end{equation}
\begin{equation}
F^{(PR)}_0(Z_f) = \frac{2\pi \alpha Z_f}{1-e^{-2\pi \alpha Z_f}}.
\end{equation}
This leads to the phase-space factor for $\beta^+$ - decay
\begin{equation}
f^{(+)}_0 \approx \frac{1}{30}(E^5_0-10E^2_0+15E_0-6)F^{(PR)}_0(-Z_f).
\end{equation}
The endpoint energy $E_0$ can be extract by using the following relation
\begin{equation}
E_0 = \frac{Q_{EC}-m_ec^2}{m_ec^2}. 
\end{equation}
The experimental $\beta$ -decay $Q_{EC}$ values are taken from \cite{ENSDF}.

Usually, $ft$ values are large so expressed in terms of `log $ft$ values'. The log$ft$ $\equiv$ log$_{10}({f_0}t_{1/2}[s])$.

  The total half-life can be calculated with the help of the  partial half-life ($t_{i}$) of the daughter state $i$ using the following expression: 
\begin{equation}
 t_{1/2}= \left({\sum_i {\frac{1}{t_i}}}\right)^{-1}.
\end{equation}

The expression for the partial half-life of the allowed $\beta$-decay is taken from \cite{suhonen}.

The branching ratio ${b_r}$ is related to partial half-life ${t_i}$ and the total half-life $t_{1/2}$ of the allowed $\beta$-decay as
\begin{equation}
 t_{i}= {\frac{t_{1/2}}{b_r}}.
 \label{br}
\end{equation}

\begin{figure*}
\begin{center}
\resizebox{0.38\textwidth}{0.32\textwidth}{\includegraphics{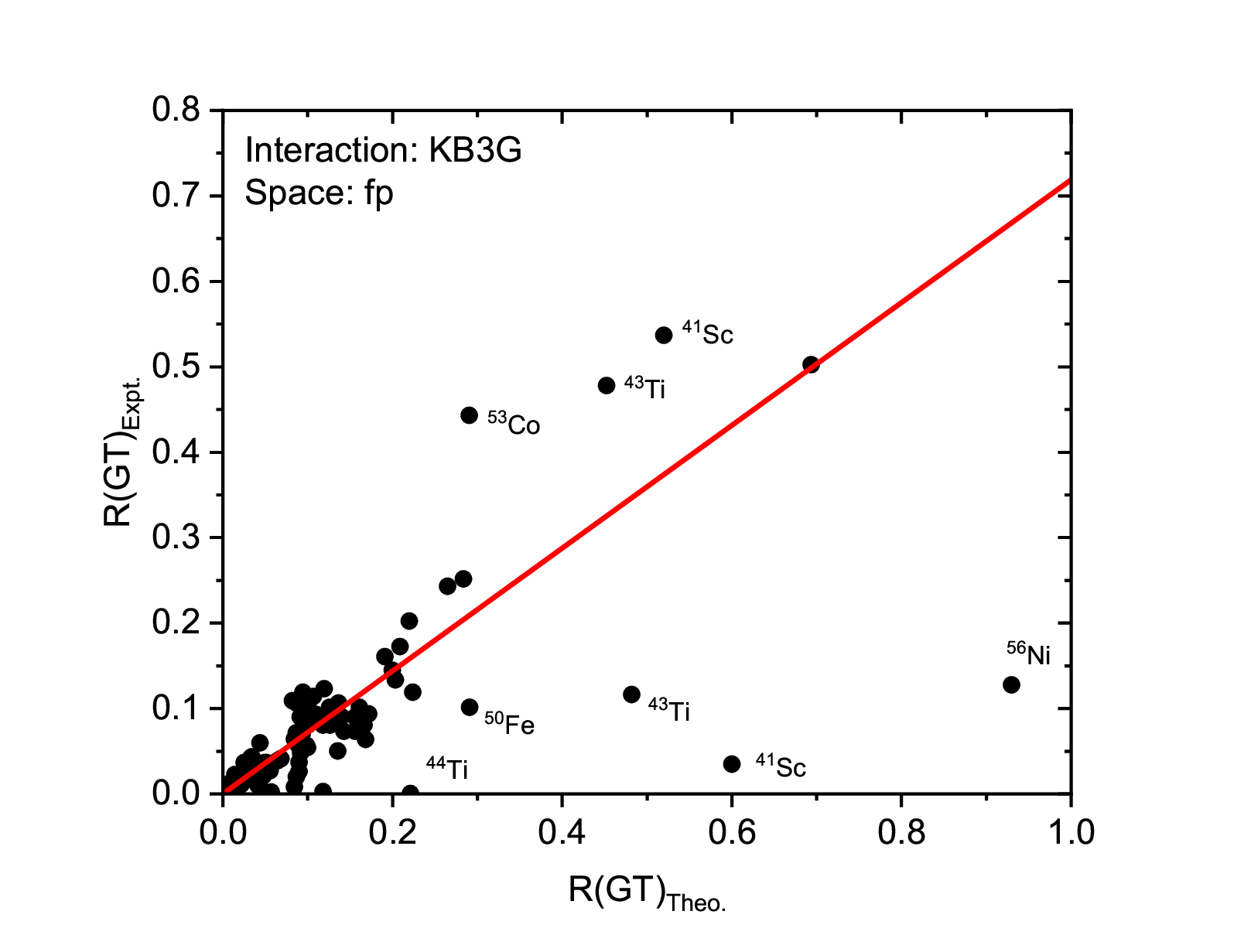}}
\resizebox{0.38\textwidth}{0.32\textwidth}{\includegraphics{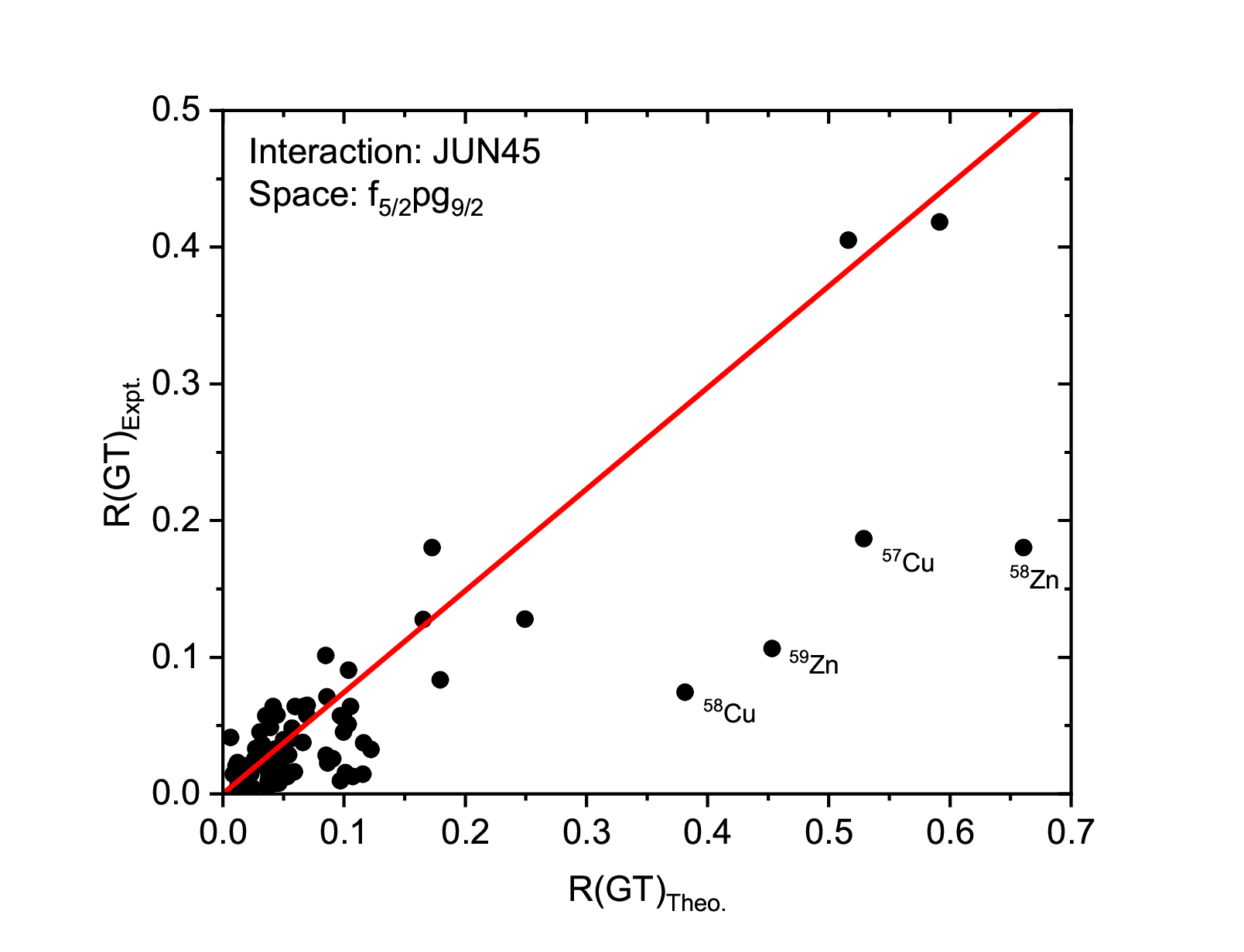}}
\caption{\label{qf} The experimental versus theoretical matrix element $R(GT)$ values are compared. SM calculations are based on the ``free-nucleon" Gamow-Teller operator. Each transition is represented by a point in the graph. The theoretical and experimental values are taken on x, y coordinates, respectively. The data points which are away from the average quenching factor are indicated by $^{A}X$.}
\end{center}
\end{figure*}

\section{Shell model Hamiltonian and quenching factor}

In the present work, for $Z = 21 -30$ nuclei we used KB3G \cite{kb3g} and JUN45 \cite{jun45} effective interactions for shell model calculations. The shell model code NuShellX@MSU \cite{MSU-NSCL} is used for the diagonalization of energy matrices.

 We have reported $\beta$-decay half-lives for nuclei in two different model spaces using two different effective interactions KB3G and JUN45. The mass dependence and original monopole changes in KB3 are known as KB3G effective interaction \cite{kb3g}. The idea behind this is to treat properly the $N=Z=28$ shell closure and its surroundings. The single-particle energies are taken to be -8.6000, -6.6000, -4.6000, and -2.1000 MeV for the $f_{7/2}$, $p_{3/2}$, $p_{1/2}$ and $f_{5/2}$ orbits, respectively for KB3G effective interaction.

Honma {\it et al} \cite{jun45} developed JUN45 effective interaction for $f_{5/2}pg_{9/2}$ model space. The JUN45 interaction is a realistic interaction that is based on Bonn-C potential. Further, this interaction is fitted by 400 experimental data (binding and excitation energies) with mass numbers A = 63 - 96. For JUN45 interaction the single-particle energies are taken to be -9.8280, -8.7087, -7.8388, and -6.2617 MeV for the $p_{3/2}$, $f_{5/2}$, $p_{1/2}$, and $g_{9/2}$ orbits, respectively. A large number of experimental data are taken for fitting around $N=50$ shell closure. 
The full-fledged shell-model calculations have been performed for all transitions.

The matrix element $M(GT)$ \cite{brownrgt} can be defined in terms of reduced transition probability $B(GT)$, by following
\begin{equation}
M(GT)= [(2J_{i}+1)B(GT)]^{1/2},
\end{equation}
where $J_{i}$ represents the total angular momentum of the initial state.
The matrix element $M(GT)$ is normalized to the ``expected" total strength W in order to get effective axial-vector coupling constant $g_{A}$, W is defined as
\begin{equation}
  W=\left\{
  \begin{array}{@{}ll@{}}
    |g_{A}/g_{V}|[(2J_{i}+1)3|N_{i}-Z_{i}|]^{1/2} , & for N_{i} \neq  Z_{i},\\
    |g_{A}/g_{V}|[(2J_{f}+1)3|N_{f}-Z_{f}|]^{1/2} , & for N_{i} = Z_{i},
  \end{array}\right.
\end{equation}

The matrix elements $R(GT)$ are defined as
\begin{equation}
R(GT) = M(GT)/W.
\end{equation}

The theoretical versus experimental $R(GT)$ values are plotted in Fig.~\ref{qf}. From table 2, the experimental log$ft$ values are used to get the $R(GT)_{Expt}$.

The theoretically calculated Gamow-Teller strengths on the basis of model independent Ikeda sum rule ``3(N-Z)" are larger than observed values, so we needed a quenching factor (q) for a particular model space. The average of all the ratios  between experimental and theoretical $R(GT)$ values give the quenching factor (q) for a given model space. The straight line in Fig.~\ref{qf} gives the average quenching factor. In the present work, we have obtained two different quenching factors: 
$q$ = 0.719 $\pm$ 0.050 for $pf$ space using KB3G interaction, and $q$ = 0.743 $\pm$ 0.030 for $f_{5/2}pg_{9/2}$ space using JUN45 interaction. Fig. 1 shows that a few data points are notably away from the straight line, we marked those data points by $^{A}X$. There are four data points in $fp$ space and eight data points in $f_{5/2}pg_{9/2}$ space which are away from the straight line. After excluding these data points we get the quenching factor $q$ = 0.669 $\pm$ 0.020 and $q$ = 0.768 $\pm$ 0.030 for $pf$ and $f_{5/2}pg_{9/2}$ space, respectively. 

\begin{table*}
\begin{center}
\caption{Computed phase space factors for $\beta^+$/EC decay (considered only those nuclei in which the EC phase space factors are significant), log$(f^{(+)}_0+f^{(EC)}_{0})$ values, together with experimental $Q$ values taken from \cite{ENSDF}.}
\begin{tabular}{cccccc}
\hline
$^{A}Z_{i}(J^{\pi})$ &  $^{A}Z_{f}(J^{\pi})$ & $Q^{(EXPT.)}$ (MeV) & $f^{(+)}_0$ & $f^{(EC)}_0$ &log$(f^{(+)}_0+f^{(EC)}_{0})$\\
\hline
$^{43}$Sc($7/2^-$)  & $^{43}$Ca($7/2_1^-$)    &  2.220 &   7.137  &  0.425   & 0.878\\
                    & $^{43}$Ca($5/2_1^-$)    &  1.848 &   1.779  &  0.294   & 0.316\\
$^{45}$Ti($7/2^-$)  & $^{45}$Sc($7/2_1^-$)   & 2.062  &  4.063    &  0.421   & 0.652  \\
                    & $^{45}$Sc($5/2_1^-$)   & 1.519  &  0.288    &  0.228   & -0.287  \\
                    & $^{45}$Sc($5/2_2^-$)   & 1.342  &  0.066    &  0.178   & -0.614  \\
                    & $^{45}$Sc($7/2_2^-$)   & 1.088  &  0.0005    &  0.116   & -0.933  \\
$^{53}$Fe($7/2^-$)&$^{53}$Mn($7/2_1^-$)      &  3.742 &  174.33    & 2.292   & 2.247 \\
                    &$^{53}$Mn($5/2_1^-$)    &  3.364 &  92.39    &  1.851   & 1.974\\
                    &$^{53}$Mn($9/2_1^-$)    & 2.122  &  4.510    &  0.734   & 0.720\\
                    &$^{53}$Mn($5/2_2^-$)    & 1.468  &  0.179    &  0.350   & -0.277 \\
 
$^{55}$Co($7/2^-$)  &$^{55}$Fe($5/2_1^-$)    & 2.521 &  14.546    & 1.161    & 1.196 \\
                    &$^{55}$Fe($7/2_1^-$)    & 2.135 &  4.581    &  0.831   &  0.733\\
                    &$^{55}$Fe($7/2_2^-$)    & 2.043 &  3.313    &  0.761   &  0.610\\
                    &$^{55}$Fe($5/2_2^-$)    & 1.308 &  0.039    &  0.310   &  -0.457\\
                    &$^{55}$Fe($9/2_1^-$)    & 1.240 &  0.016    &  0.279   &  -0.530\\
                    &$^{55}$Fe($9/2_1^-$)    & 1.151 &  0.003    &  0.240   &  -0.615\\
$^{57}$Co($7/2^-$)&$^{57}$Fe($5/2_1^-$)    & 0.699  &  --    &  0.087   & -1.060 \\
$^{58}$Co($2^+$)&$^{58}$Fe($2_1^+$)        &  1.497 &  0.216   &   0.407  & -0.206 \\
                &$^{58}$Fe($2_2^+$)        &  0.633 &  --    &  0.071  & -1.148 \\

$^{56}$Ni($0^+$)&$^{56}$Co($1^+$)          &  0.415 &   --   &  0.034   &  -1.468\\
$^{57}$Ni($3/2^-$)&$^{57}$Co($3/2_1^-$)    & 1.886  &   1.736   &  0.722   &0.391 \\
                  &$^{57}$Co($1/2_1^-$)    & 1.759  &   0.976   &  0.628   &0.205 \\
                  &$^{57}$Co($3/2_2^-$)    & 1.506  &   0.224   &  0.459   &-0.166 \\
                  &$^{57}$Co($5/2_1^-$)    & 1.344  &   0.057   &  0.365   & -0.375\\
$^{61}$Cu($3/2^-$)&$^{61}$Ni($3/2_1^-$)     &  2.238 &  6.087    &  1.131   & 0.858\\
                    &$^{61}$Ni($5/2_1^-$)   &  2.171 &  4.900    &  1.064   & 0.775\\
                    &$^{61}$Ni($1/2_1^-$)   &  1.955 &  2.241    &  0.862   & 0.491\\
                    &$^{61}$Ni($1/2_2^-$)   &  1.582 &  0.359    &  0.563   & -0.035\\
                    &$^{61}$Ni($5/2_2^-$)   &  1.329 &  0.048    &  0.396   & -0.353\\
                    &$^{61}$Ni($3/2_2^-$)   &  1.138 &  0.002    &  0.290   & -0.535\\
                    &$^{61}$Ni($5/2_3^-$)   &  1.105 &  0.0007    &  0.273   & -0.563\\
                    &$^{61}$Ni($3/2_3^-$)   &  1.052 &  0.00003    &  0.247   &-0.607 \\

$^{64}$Cu($1^+$)&$^{64}$Ni($0_1^+$)       &  1.675&   0.617   &  0.631   & 0.096\\
$^{60}$Zn($0^+$)&$^{60}$Cu($1_1^+$)           & 4.109  &   269.56   &  4.240   &2.437 \\
                    &$^{60}$Cu($1_2^+$)       & 3.806  &   172.51   &  3.636   & 2.245\\
                    &$^{60}$Cu($1_3^+$)       & 3.501  &   105.18   &  3.075   & 2.034\\

$^{62}$Zn($0^+$)&$^{62}$Cu($1_1^+$)           & 1.620  &  0.439    &  0.653   & 0.038\\ 
                    &$^{62}$Cu($1_2^+$)       & 1.072  &  0.0002    & 0.284    & -0.547\\ 
                    &$^{62}$Cu($1_3^+$)       & 0.983  &  --    & 0.238    & -0.623\\ 

$^{65}$Zn($5/2^-$)&$^{65}$Cu($3/2_1^-$)      & 1.352  &  0.059    &  0.453   & -0.291\\                  
\hline 
\end{tabular}
\end{center}
\end{table*}

\begin{table*}
\begin{center}
\caption{The theoretical excitation energies, log$ft$ values, and branching
ratios of $\beta^+$-decays of the concerned nuclei are compared with the experimental values. Where the experimental ground state energy and parity $J^\pi$ of the parent nucleus is uncertain is indicated by an asterisk. In the last column, the references of the experimental data are given.}
\begin{tabular}{r@{\hspace{4pt}}c@{\hspace{2pt}}c@{\hspace{2pt}}c@{\hspace{2pt}}c@{\hspace{2pt}} c@{\hspace{2pt}}c@{\hspace{2pt}}c@{\hspace{2pt}}c@{\hspace{2pt}}c@{\hspace{2pt}}c@{\hspace{2pt}} c@{\hspace{2pt}}c@{\hspace{2pt}}c@{}}
\hline
&    &&\multicolumn {2} {c} {Ex. energy (keV)} &&\multicolumn {2} {c} {log$ft$ value}&& \multicolumn {2} {c} {Branching ($\%$)}\\
\cline{4-5}
\cline{7-8}
\cline{10-11}
$^{A}$Z$_{i}(J^{\pi})$  & $^{A}$Z$_{f}(J^{\pi})$  &&
      \multicolumn{1}{c}{Theo.}&\multicolumn{1}{c}{Expt.} &&
      \multicolumn{1}{c}{Theo.}& \multicolumn{1}{c}{Expt.} &&
      \multicolumn{1}{c}{Theo.}&\multicolumn{1}{c}{Expt.}&&
      Ref.\\
\hline
$^{41}$Sc(${7/2}^-$)& $^{41}$Ca($7/2^-$) && 0     & 0    && 3.768  &3.452   && 99.97   &   99.96 &&  \cite{41sc}\\
                   & $^{41}$Ca($5/2^-$)  &&  6500 & 2575 &&  3.643 &5.830   &&  0.027    &   0.023 &&              \\ 
$^{42}$Sc($0^+$)   & $^{42}$Ca($1^+$)    && 9068  & --   &&  3.969 &  --    &&   37.32   &--       &&  \\
$^{43}$Sc($7/2^-$) & $^{43}$Ca($7/2^-$)  && 0     &  0   &&  5.240 & 5.04   &&  69.67    & 77.54    && \cite{ENSDF} \\
                   & $^{43}$Ca($5/2^-$)  && 216   &  373 &&  5.113 & 4.98   &&  25.57    & 22.53    &&  \\
                   &                     && 3425  &  1931&&  5.527 & 5.68   &&  4.75     &0.02      &&  \\
$^{42}$Ti($0^+$)   & $^{42}$Sc($1^+$)    && 342   & 611  &&  3.216 &  3.495     &&  80.23   &56.06   && \cite{42ti} \\
                   & $^{42}$Sc($1^+$)    && 4111  & 1888 &&  4.694 &  4.80      &&  0.03    &0.41 && \\ 
$^{43}$Ti($7/2^-$) & $^{43}$Sc($7/2^-$)  && 0     &  0   &&  3.888 & 3.554        && 88.03  & 90.28    && \cite{43ti} \\
                   &                     && 3115  & 1408 &&  6.145 & 5.130        && 0.015   & 0.67    &&  \\
                   & $^{43}$Sc($5/2^-$)  && 2162  &845   &&  3.833 & 4.78         && 11.93  & 2.6    &&  \\
                   &                     && 3740  & 2288 &&  6.763 & 3.85         && 0.001  & 4.6    &&  \\
                   &                     && 4309  &  2335&&  5.589 & 4.91         && 0.004  & 0.38    &&  \\
                   
$^{45}$Ti($7/2^-$) & $^{45}$Sc($7/2^-$)  && 0  &  0     && 4.598  & 4.591           && 89.94 & 99.69    && \cite{45ti} \\
                   &                     && 2519  &  1408  && 5.777  & 5.78         &&  0.12 & 0.09   &&  \\
                   & $^{45}$Sc($5/2^-$)  && 1533  &  720   && 6.064  & 6.26         &&  0.35 & 0.14    &&  \\
                   & $^{45}$Sc($9/2^-$)  && 1727  &  1662  && 5.275  & 5.55         &&  0.001 & 0.06    &&  \\
$^{44}$V($(2)^{*+}$)& $^{44}$Ti($2^+$)&&  1300    & 1083  &&  4.497 &  4.70  &&  70.21   &32  && \cite{44v} \\
                    & $^{44}$Ti($2^+$)&& 3360     & 2530  &&  4.628 &  4.56  &&  21.72    &23 && \\ 
$^{45}$V($7/2^-$) & $^{45}$Ti($7/2^-$)&&0         &0      && 4.305  &3.64    && 89.89  & 95.6    && \cite{45v} \\
                  & $^{45}$Ti($5/2^-$)&&12        &40     && 5.701  &5.0     && 3.57   &4.3     &&         \\    
$^{46}$V($0^+$)   & $^{46}$Ti($1^+$)  && 3809     & 4315  &&  4.953 &  5.0   &&  6.62  &0.01  && \cite{46v} \\
$^{47}$V($3/2^-$) & $^{47}$Ti($5/2^-$)&& 0        &0      &&4.786   &4.901   &&  99.67  & 99.55    && \cite{47v} \\
                  &                   && 1950     & 2166  &&  5.836 &6.25    &&  0.02  & 0.01    &&       \\
                  & $^{47}$Ti($3/2^-$)&& 1338     & 1550  &&  6.576 &6.08     && 0.02   & 0.04 &&         \\
                  &                   && 1922     & 2163  &&  5.294 &5.36     && 0.10   &  0.07&&         \\
                  &                   && 2294     & 2548  &&  5.913 &5.77     && 0.01   &  0.01 &&         \\
                  & $^{47}$Ti($1/2^-$)&& 1848     & 1794  && 5.195  &  5.10   && 0.15   &  0.28 &&         \\
                  &                   && 2844     & 2793  && 4.812  &  5.18   && 0.002   & 0.003 &&         \\
$^{45}$Cr(($7/2^-$)*) & $^{45}$V($7/2^-$) && 0  & 0     && 4.599  & --   &&  40.06 & --  && \cite{45cr} \\
                   &                   && 2425  & 4800  && 5.250  & 3.68   &&  2.83  & 19.6  &&         \\
                   & $^{45}$V($9/2^-$) && 1565  & 1322  && 4.685  & -- &&  16.11 & --  &&         \\
$^{47}$Cr($3/2^-$) & $^{47}$V($3/2^-$) && 0  & 0     && 4.535  & 3.70   &&  76.11 & 96.1  && \cite{47cr} \\
                   & $^{47}$V($5/2^-$) && 3  & 87  && 5.056  & 5.1 &&  22.88 & 3.9  &&         \\
$^{48}$Mn($4^+$)   &$^{48}$Cr($4^+$)   && 2700   & 1858 &&  5.665 &   5.4     &&  8.540  &  5.9   && \cite{48mn}\\
                   &                   && 4932   & 4428 &&  4.352 &   4.6     &&  68.32  &  10.0  &&  \\
                   &                   && 5394   & 5792 &&  7.032 &   3.49     &&  0.11  &  58.3     &&  \\
$^{49}$Mn($5/2^-$) &$^{49}$Cr($5/2^-$) && 0      & 0   && 4.388  &  3.68      &&  86.42  &   91.8  && \cite{ENSDF}\\
                   &$^{49}$Cr($7/2^-$) && 280    & 272 && 5.109  &   4.8  &&  13.43  &  5.8  &&  \\
                   &                   && 2299   & 2504 &&  6.324 &   4.3     &&  0.14  &    2.3   &&  \\
$^{50}$Mn($0^+$)   &$^{50}$Cr($1^+$)   &&  3539  & 3628 && 5.004  &   5.14    &&  6.41  &   0.056  && \cite{46v}\\
                   &                   &&  4814  & 4998 && 6.254  &   5.90 &&  0.13  &  0.0007  &&  \\
$^{51}$Mn($5/2^-$) &$^{51}$Cr($7/2^-$) &&  0     &  0     && 5.315  &   5.297     &&  99.73  &  99.62   && \cite{51mn}\\
                   &                   &&  1307  &  1557  && 7.099  &   7.09     &&   0.06 &  0.007   && \\
                   &                   &&  2321  &  2312  && 5.678  &   5.662     &&  0.05  &  0.09  && \\
                   &$^{51}$Cr($5/2^-$) && 1168   & 1353 && 7.178  &  7.32   &&  0.06  &   0.05 &&  \\
                   &                   && 1971   & 2001 &&  6.214 &  6.314   && 0.05   &  0.03  &&  \\
$^{48}$Fe($0^+$)   &$^{48}$Mn($1^+$)   && 90     & 403  && 4.085  &    3.9    &&  33.75 &  42   && \cite{48fe}\\
                   &                   && 1702   & 3204 && 5.033  &    4.8    &&  2.10  &  1.0  &&  \\
                   &                   && 2123   & 3495 && 4.056  &    4.5    &&  16.9 &  1.8  &&  \\
                   &                   && 2823   & 3619 && 4.197  &    4.7    &&  9.13  &  0.9  &&  \\
                   &                   && 3187   & 3713 && 6.974  &    4.6    &&  0.01  &  1.3  &&  \\
                   &                   && 3291   & 4299 && 4.986  &    4.4    &&  1.20  &  1.2  &&  \\
                   &                   && 3523   & 4399 && 4.626  &    4.5    &&  2.49  &  0.9  &&  \\
                   &                   && 3728   & 4517 && 4.404  &    4.3    && 3.77 &  1.3  &&  \\
                   &                   && 3931   & 4755 && 4.374  &    4.4    &&  3.68 &  0.8  &&  \\
$^{49}$Fe(($7/2^-$)*) &$^{49}$Mn($5/2^-$) && 0   &  0     && 5.985  &   --        &&  4.79    &  --    && \cite{49fe}\\
                   &                   && 1873   &  3959  && 5.349  &   5.97      &&  9.10   &  1.2   && \\
                   
\hline 
\end{tabular}
\end{center}
\end{table*}

\addtocounter{table}{-1}

\begin{table*}
  \begin{center}
    \leavevmode
    \caption{{\em Continuation.\/}}
\begin{tabular}{r@{\hspace{4pt}}c@{\hspace{2pt}}c@{\hspace{2pt}}c@{\hspace{2pt}}c@{\hspace{2pt}} c@{\hspace{2pt}}c@{\hspace{2pt}}c@{\hspace{2pt}}c@{\hspace{2pt}}c@{\hspace{2pt}}c@{\hspace{2pt}} c@{\hspace{2pt}}c@{\hspace{2pt}}c@{}}
\hline
&    &     & \multicolumn {2} {c} {Ex. energy (keV)} &&\multicolumn {2} {c} {log$ft$ value}&& \multicolumn {2} {c} {Branching ($\%$)} & \\
\cline{4-5}
\cline{7-8}
\cline{10-11}
$^{A}$Z$_{i}(J^{\pi})$  & $^{A}$Z$_{f}(J^{\pi})$  &&
      \multicolumn{1}{c}{Theo.}&\multicolumn{1}{c}{Expt.} &&
      \multicolumn{1}{c}{Theo.}& \multicolumn{1}{c}{Expt.} &&
      \multicolumn{1}{c}{Theo.}&\multicolumn{1}{c}{Expt.}&& Ref.\\
\hline

                   &$^{49}$Mn($7/2^-$) && 279   & 261  && 4.900  &  --   &&   52.02 &   -- &&  \\
                   &                   && 2298   & 4381 &&7.576   &  5.83   &&  0.04  &  1.4  &&  \\
                   &                   && 2525   & 4814 &&6.923   &  4.36   &&  0.18  &  34.5  &&  \\  
$^{50}$Fe($0^+$) & $^{50}$Mn($1^+$) && 426   &  651     && 3.967  &  3.81  &&  39.63 &  22.5 && \cite{50fe} \\
                 &                  && 2415  &  2403    && 4.352  &  4.36  &&  3.53 &  1.47 &&                \\
                 &                  && 3119  &  2684    && 4.543  &  4.56  &&  1.15 &  0.70 &&                \\
                 &                  && 3302  &  3380    && 4.401  &  4.14  &&  1.32 &  0.84 &&                \\
                 &                  && 3539  &  3643    && 5.002  &  4.80  &&  0.25 &  0.15 &&                \\
                 &                  && 3787  &  4012    && 5.245  &  5.10  &&  0.10 &  0.04 &&                \\
                 &                  && 3898  &  4316    && 4.114  &  4.60  &&  1.28 &  0.08 &&                \\
$^{51}$Fe($5/2^-$) &$^{51}$Mn($5/2^-$)&& 0      & 0    && 4.323  &  3.653  &&  82.69 &  93.7   && \cite{51fe}\\
                   &$^{51}$Mn($7/2^-$)&& 265    & 237  && 5.074  &  4.86   &&  12.26 &  5.0    &&  \\
                   &$^{51}$Mn($3/2^-$)&& 1854   & 1825 && 5.627  &  5.32   &&  0.98  &  0.49   &&  \\
                   &                  && 2063   & 2140 && 5.323  &  5.51   &&  1.63  &  0.24   &&  \\
                   &                  && 2943   & 2914 && 5.189  &  5.54   &&  1.50  &  0.10   &&  \\
                   &                  && 3301   & 3555 && 5.185  &  5.00   &&  0.91  &  0.16   &&  \\
$^{53}$Fe($7/2^-$) &$^{53}$Mn($7/2^-$)&& 0      & 0    && 5.132  &  5.22   && 72.27     &   55.95  && \cite{53fe}\\
                   &                  && 2409   & 2685 && 7.877  &  5.10   && 0.0002   &   0.01   &&  \\
                   &$^{53}$Mn($5/2^-$)&& 385     & 378  && 5.276  & 5.06    && 26.84   &  42.09  &&  \\
                   &                  && 2055   & 2273 && 6.321  & 4.90    && 0.03   &  0.38   &&  \\
                   &                  && 2937   & 3126 && 6.515  & 4.50    && 0.002   &  0.14     &&  \\
                   &$^{53}$Mn($9/2^-$)&& 1815   & 1619 && 6.974  &  5.30   && 0.01     &  1.0   &&  \\
                   &                  && 3007   & 2946 && 6.673  &  5.10   &&  0.002  &  0.05    &&  \\
                   &                  && 3465   & 3248 && 5.933  &  4.80   &&  0.001  & 0.04     &&  \\
$^{50}$Co(($6^+$)*)&$^{50}$Fe($6^+$)  && 3151   & 3159 && 7.032  & 4.78   &&  0.87   & 15    && \cite{50co}\\
                   &                  && 3739   & 8458 && 4.889  & 3.32   &&  99.12  & 42.1    &&  \\
$^{52}$Co(($6^+$)*)&$^{52}$Fe($6^+$)   && 4394   & 4326 && 4.331  &  --  &&  58.02  &  --   && \cite{52co}\\
                   &                   && 4936   & 5655 && 4.355  &  3.4  && 41.97  &  --   &&  \\
$^{53}$Co(($7/2^-$)*) &$^{53}$Fe($7/2^-$)&& 0   & 0    && 4.266  &  3.62   &&  87.42  &  94.49   && \cite{53co}\\
                   &$^{53}$Fe($9/2^-$)&& 1459   & 1328 && 4.652  &  4.44   &&  12.57   &  5.6    &&  \\
$^{55}$Co($7/2^-$) &$^{55}$Fe($5/2^-$)&& 946   & 931  && 5.923  &  6.25   && 67.22   &  51.6   && \cite{ENSDF}\\
                   &                  && 2110   & 2144 && 6.346  &  6.69   && 0.56   &  0.55   &&  \\
                   &                  && 2688   & 2578 && 7.877  &  7.44   && 0.05   &  0.04   &&  \\
                   &$^{55}$Fe($7/2^-$)&& 1398   & 1316 && 6.114  &  6.725   && 14.95   &  5.6   &&  \\
                   &                  && 1692   & 1408 && 5.979  &  5.785   && 15.31   &  36.3   &&  \\
                   &$^{55}$Fe($9/2^-$)&& 2314   & 2212 && 5.937  &  6.110   && 1.22   &  1.86   &&  \\
                   &                  && 2592   & 2301 && 6.309  &  5.780   && 0.17   &  3.4   &&  \\  
$^{57}$Co($7/2^-$) &$^{57}$Fe($5/2^-$)&& 0     & 136 && 6.286  &  6.450   &&  77.25  &  99.80   && \cite{ENSDF}\\
                   &                  && 981   & 706 && 7.877  &  7.70    &&  22.74  &   0.17  &&  \\

$^{58}$Co($2^+$)   &$^{58}$Fe($2^+$)   && 802   & 810  && 6.386  & 6.61   && 98.19  &  98.8   && \cite{ENSDF}\\
                   &                   && 1789   & 1674 && 7.178 & 7.69   && 1.80   & 1.21      &&  \\
$^{54}$Ni($0^+$)   &$^{54}$Co($1^+$)   && 871    & 936  && 3.899  & 3.84   &&  46.07 &  19.8   && \cite{54ni}\\
                   &                   && 2075   & 2424 && 5.397  & 5.43   &&  0.66 &  0.16   &&  \\
                   &                   && 2715   & 3376 && 5.555  & 4.67   &&  0.28  &  0.37   &&  \\
                   &                   && 3934   & 3889 && 5.226  & 4.43   &&  0.21  &  0.37   &&  \\
                   &                   && 4547   & 4293 && 5.250  & 4.46   &&  0.10  &  0.21   &&  \\
                   &                   && 4824   & 4323 && 4.500  & 4.70   &&  0.42  &  0.11   &&  \\
                   &                   && 4993   & 4543 && 4.912  & 4.46   &&  0.13  &  0.15   &&  \\
                   &                   && 5095   & 4822 && 3.994  & 4.37   &&  0.97  &  0.12   &&  \\
                   &                   && 5323   & 5202 && 7.275  & 4.88   &&  0.0003  &  0.02   &&  \\
$^{56}$Ni($0^+$)   &$^{56}$Co($1^+$)   && 1299   & 1720 && 4.308  &  4.40  &&  100  &  100   && \cite{56ni}\\
$^{57}$Ni($3/2^-$) &$^{57}$Co($3/2^-$)&& 1815   & 1377 && 5.583  &  5.64   &&  75.90  &  64.5   && \cite{ENSDF}\\
                   &                  && 2043   & 1757 && 6.372  &  6.22   &&  3.42   &  5.66   &&  \\
                   &$^{57}$Co($1/2^-$)&& 1967   & 1504 && 5.991  &  6.05   &&  19.34   &  17.04   &&  \\
                   &$^{57}$Co($5/2^-$)&& 1879   & 1919 && 6.576  &  5.74   &&  3.42   &  12.3   &&  \\
                   &                  && 2356   & 2133 && 6.415  &  8.13   &&  0.001  &  0.03    &&  \\

 $^{57}$Cu($3/2^-$) &$^{57}$Ni($3/2^-$)&& 0   & 0    && 3.655  &  3.67   &&  83.36  &   89.9  && \cite{57cu}\\
                   &$^{57}$Ni($5/2^-$)&&  1119  & 768  && 8.877  &  5.44   &&  0.0002  &   0.94  &&  \\
                   &$^{57}$Ni($1/2^-$)&& 1989   & 1112 && 3.752  & 4.37    &&  16.63  &  8.6   &&  \\ 
$^{58}$Cu($1^+$)   &$^{58}$Ni($0^+$)   && 0   & 0    && 3.735  & 4.87   && 96.39   &  81.2   && \cite{58cu}\\
                   &                   && 3004   & 2943 && 4.391  & 4.77   && 1.99   &  10.1   &&  \\                   
 \hline
\end{tabular}
\end{center}
\end{table*}

\addtocounter{table}{-1}

\begin{table*}
  \begin{center}
    \leavevmode
    \caption{{\em Continuation.\/}}
\begin{tabular}{r@{\hspace{4pt}}c@{\hspace{2pt}}c@{\hspace{2pt}}c@{\hspace{2pt}}c@{\hspace{2pt}} c@{\hspace{2pt}}c@{\hspace{2pt}}c@{\hspace{2pt}}c@{\hspace{2pt}}c@{\hspace{2pt}}c@{\hspace{2pt}} c@{\hspace{2pt}}c@{\hspace{2pt}}c@{}}
\hline
&    &     & \multicolumn {2} {c} {Ex. energy (keV)} &&\multicolumn {2} {c} {log$ft$ value}&& \multicolumn {2} {c} {Branching ($\%$)} & \\
\cline{4-5}
\cline{7-8}
\cline{10-11}
$^{A}$Z$_{i}(J^{\pi})$  & $^{A}$Z$_{f}(J^{\pi})$  &&
      \multicolumn{1}{c}{Theo.}&\multicolumn{1}{c}{Expt.} &&
      \multicolumn{1}{c}{Theo.}& \multicolumn{1}{c}{Expt.} &&
      \multicolumn{1}{c}{Theo.}&\multicolumn{1}{c}{Expt.}&& Ref.\\
\hline
                   &                   && 4296   & 3532 && 4.926  & 6.64   && 0.004   &  0.07   &&  \\
                   &$^{58}$Ni($2^+$)   && 1298   & 1454 && 5.728  & 6.20   && 0.41   &  1.4   &&  \\
                   &                   && 2969   & 2776 && 4.837  & $>$6.4 && 0.74   &  $<$0.28   &&  \\
                   &                   && 3856   & 3038 && 4.888  & 6.23   && 0.24   &  0.32   &&  \\
                   &$^{58}$Ni($1^+$)   && 3309   & 2902 && 7.877  & 5.13   && 0.001   &  4.6   &&  \\                                         
$^{59}$Cu($3/2^-$) &$^{59}$Ni($3/2^-$)&& 0      &0     && 5.168  & 5.0    && 64.04   &  58.05   && \cite{59cu}\\
                   &                  && 1169   &878   && 5.154  & 5.3    && 12.69   &  8.78    &&  \\
                   &                  && 2603   &1734  && 5.963  & 5.3    && 0.08     &  1.99   &&  \\
                   &                  && 2950   &2414  && 5.171  & 5.5    && 0.17    &  0.20   &&  \\
                   &$^{59}$Ni($5/2^-$)&& 542    &339   && 6.169  & 5.8    && 3.17   &  5.89   &&  \\
                   &                  && 1644   &1188  && 6.057  & 7.0    && 0.66   &  0.11   &&  \\
                   &                  && 2312   &1679  && 6.099  & 5.4    && 0.13   &  1.61   &&  \\
                   &                  && 3373   &2681  && 5.731  & 6.0    && 0.02   &  0.03   &&  \\
                   &$^{59}$Ni($1/2^-$)&& 672    &464   && 7.032  & 6.0    && 0.36   &  3.39   &&  \\
                   &                  && 1867   &1301  && 4.406  & 4.7    && 18.67   &  19.5   &&  \\
$^{60}$Cu($2^+$)   &$^{60}$Ni($2^+$)   && 1635   & 1332 && 6.129  & 7.3   && 10.27   &  5.0    && \cite{60cu}\\
                   &                   && 2141   & 2158 && 5.442  & 6.4   && 30.90   &  15.3   &&  \\
                   &                   && 3078   & 3124 && 4.925  & 5.1   && 36.04   &  52.34   &&  \\
                   &                   && 3456   & 3269 && 5.813  & 6.0   && 2.85    &  4.99   &&  \\
                   &$^{60}$Ni($3^+$)   && 2505   & 2626 && 5.567  & 6.8   && 15.88   &  2.8    && \\
$^{61}$Cu($3/2^-$) &$^{61}$Ni($3/2^-$)&& 80   & 0    && 5.207  & 5.07    && 45.10  &  67   && \cite{61cu}\\
                   &                  && 1290   & 1099 && 5.612  & 5.86    && 0.71   &  0.68   &&  \\
                   &                  && 1840   & 1185 && 6.114  & 5.00    &&0.19   &  4.2   &&  \\
                   &$^{61}$Ni($5/2^-$)&& 0   & 67   && 6.129  & 6.35    &&  4.46  &  2.9   &&  \\
                   &                  && 1277   & 908  && 6.555  & 5.72    &&  0.12  &  1.30   &&  \\
                   &                  && 1287   & 1132 && 8.877  & 6.49    && 0.003  &  0.15    &&  \\
                   &$^{61}$Ni($1/2^-$)&& 592   & 283 && 4.860  &  5.53   && 43.05   &   8.1  &&  \\
                   &                  && 1523   & 656 && 4.725  &  4.95   && 5.33   &   13.3  &&  \\
$^{62}$Cu($1^+$)   &$^{62}$Ni($0^+$)   && 0      & 0    && 4.958  & 5.158  &&  99.30  &  99.59   && \cite{62cu}\\
                   &                   && 2192   & 2048 && 6.731  & 6.00   &&  0.01  &  0.08   &&  \\
                   &$^{62}$Ni($2^+$)   && 1820   & 1172 && 5.459  & 7.03   &&  0.62  &  0.14   &&  \\
                   &                   && 2445   & 2301 && 5.974  & 5.98   &&  0.03  &  0.02   &&  \\
$^{64}$Cu($1^+$)   &$^{64}$Ni($0^+$)   && 0   & 0    &&5.004   & 4.969   && 93.46  &  99.23   && \cite{64cu}\\
                   &$^{64}$Ni($2^+$)   && 1637   & 1345 &&5.805   & 5.504   &&  0.003  &  0.76    &&  \\
$^{58}$Zn($0^+$)   &$^{58}$Cu($1^+$)   && 0      & 0    && 3.258  & 4.1   && 80.36  &  18   && \cite{58zn}\\
                   &                   && 1476   & 1051 && 4.424  & 4.1   && 2.19    &  10   &&  \\    
$^{59}$Zn($3/2^-$) &$^{59}$Cu($3/2^-$)&& 0      & 0    && 3.774  & 3.698    && 61.71   &  94.1   && \cite{59zn}\\
                   &                  && 3336   & 4773 && 5.657  & 6.5      &&  0.07  &  0.002   &&  \\
                   &$^{59}$Cu($1/2^-$)&& 422    & 491  && 3.887  &  4.86   && 36.95   &  4.8   &&  \\
                   &                  && 3416   & 4347 && 5.285  &  6.09   && 0.15   &  0.01   &&  \\
                   &$^{59}$Cu($5/2^-$)&& 1655   & 914  && 5.899  &  5.39   &&  0.16  &  1.1   &&  \\
$^{60}$Zn($0^+$)   &$^{60}$Cu($1^+$)   && 476   & 62  && 4.901  & 5.3   && 55.53   &  20.4   && \cite{60zn}\\
                   &                   && 549   & 364 && 5.023  & 5.9   && 25.76   &  3.1    &&  \\
                   &                   && 971   & 670 && 4.461  & 4.4   && 18.64   &  72.8   &&  \\
$^{61}$Zn($3/2^-$) &$^{61}$Cu($3/2^-$)&& 0      & 0    && 5.207  &  5.40   && 77.57   &  66.7  && \cite{61zn}\\
                   &                  && 1650   & 1660 && 5.647  &  5.30   && 3.82   &  10.80   &&  \\
                   &                  && 2024   & 1932 && 6.071  &  6.25   && 0.79   &  0.79   &&  \\
                   &                  && 2379   & 2358 && 5.701  & 5.70    && 0.98   &  1.29   &&  \\
                   &                  && 2605   & 2472 && 6.224  & 5.60    && 0.19   &  1.29   &&  \\
                   &$^{61}$Cu($5/2^-$)&& 1397   & 970  && 5.923  &  7.10   &&  2.91  &   0.46  &&  \\
                   &                  && 1669   & 1394 && 6.038  & 6.69    &&  1.51  &  0.66   &&  \\
                   &                  && 1952   & 1904 && 5.832  & 6.80    &&   1.55 &   0.23  &&  \\
$^{62}$Zn($0^+$)   &$^{62}$Cu($1^+$)   && 159   & 0 && 5.212  & 4.99   &&  61.0  &  40.2   && \cite{62zn}\\    
$^{63}$Zn($3/2^-$) &$^{63}$Cu($3/2^-$)&& 0      &0     && 5.446  &  5.40   && 96.13   &  84.04   && \cite{ENSDF}\\
                   &                  && 1511   &1547  && 5.918  &  6.70   && 0.77   &  0.04   &&  \\
                   &$^{63}$Cu($1/2^-$)&& 824   & 669 && 6.647  &  5.82   && 1.02   &  7.92  &&  \\
                   
                   &$^{63}$Cu($5/2^-$)&& 1635   & 962 && 5.877  &  5.61   && 0.60   &  6.08   &&  \\
                   &                  && 1875   & 1412 && 6.297  & 5.87    && 0.12   &    0.7 &&  \\
$^{65}$Zn($5/2^-$) &$^{65}$Cu($3/2^-$)&&  0     & 0    && 5.895  &  7.450   && 28.04   &  49.96   && \cite{ENSDF}\\
                   &$^{65}$Cu($5/2^-$)&&  1569  & 1115 && 5.777  &  5.893   &&  71.95  &  50.04   &&  \\                
\hline
\end{tabular}
\end{center}
\end{table*}
\begin{figure*}
\begin{center}
\resizebox{0.88\textwidth}{0.88\textwidth}{\includegraphics{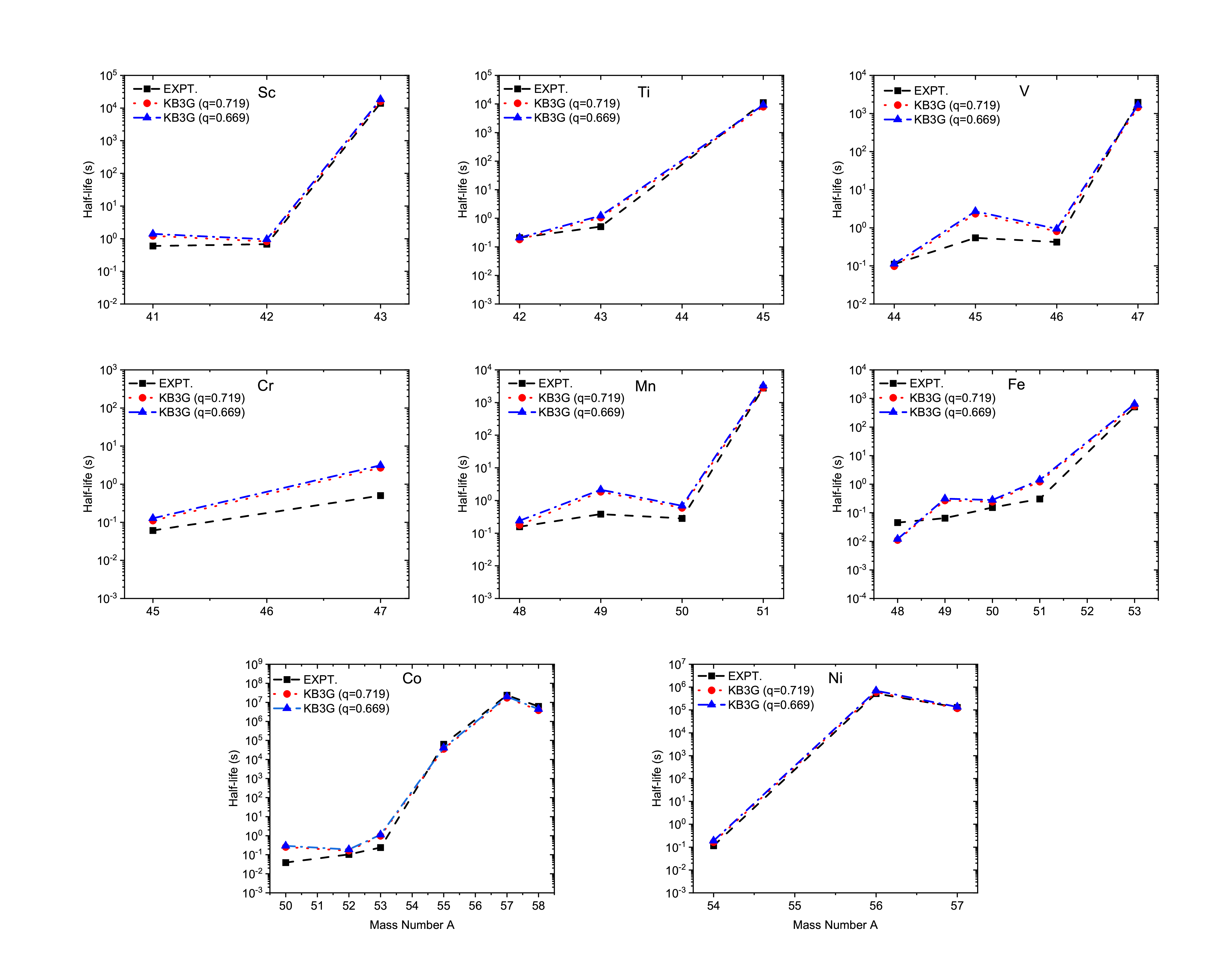}}
\caption{\label{fig1} The theoretical and experimental $\beta^+$/EC- decay half-life versus mass number A of the concerned nuclei for $fp$ space.}
\end{center}
\end{figure*}

\begin{figure*}
\begin{center}
\resizebox{0.65\textwidth}{0.30\textwidth}{\includegraphics{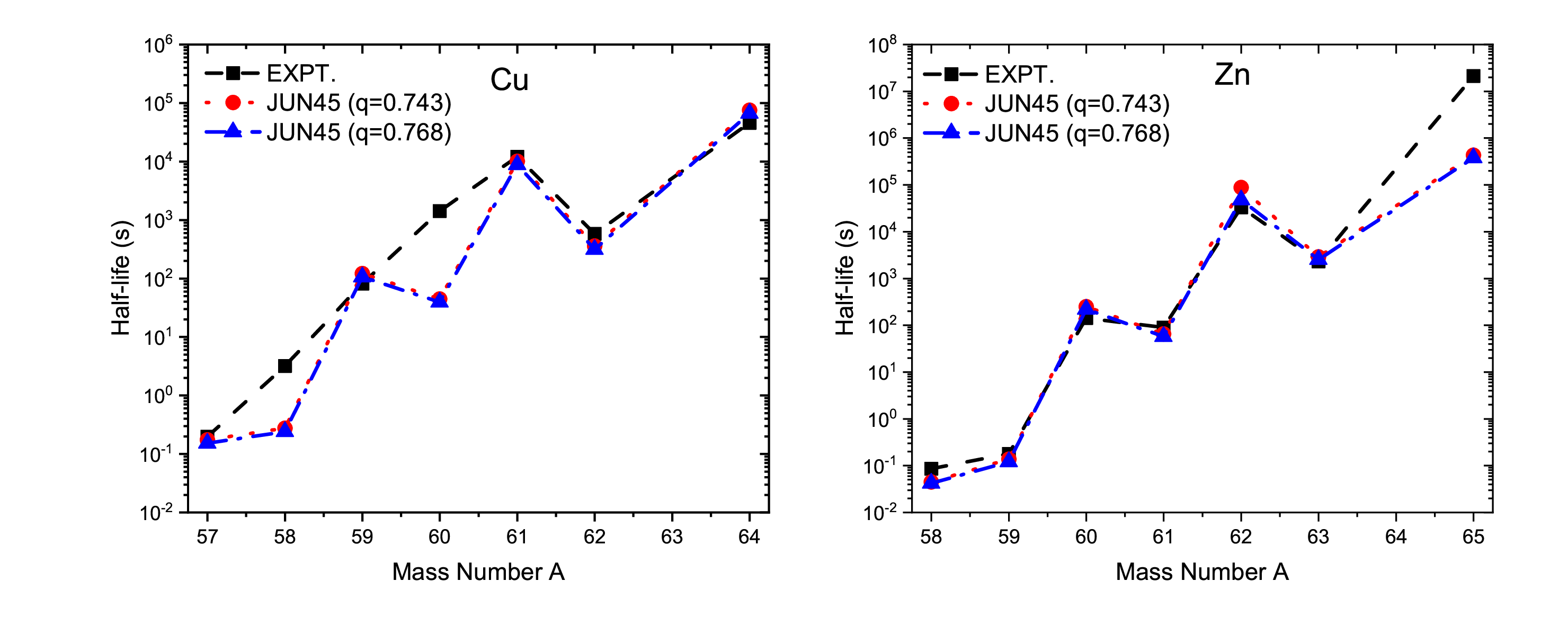}}
\caption{\label{fig11} The theoretical and experimental $\beta^+$/EC- decay half-life versus mass number A of the concerned nuclei for $f_{5/2}pg_{9/2}$ space.}
\end{center}
\end{figure*}

\begin{table*}
\begin{center}
    \leavevmode
    \caption{List of superallowed transitions with $0^+$ $\rightarrow$ $0^+$. The theoretical log$ft$ values and branching ratios of $\beta^+$/EC decay of the concerned nuclei are compared with the experimental values. The experimental ground state energy and parity $J^\pi$ of the parent and daughter nucleus are listed along with $Q$ values. In the last column the references of the experimental data are given. 
 $J_{P}^{\pi}$ ( $J_{D}^{\pi}$) and $E_P$ ($E_D$) are spin-parity and excitation energy of parent (daughter) nuclei, respectively.  }
\begin{tabular}{lrccccccccccccc}
\hline
  & & & & & & &\multicolumn{2}{c}{log$ft$}& &\multicolumn{2}{c}{Branch (\%)} & \\
\cline{8-9}
\cline{11-12}
Nuclide & Decay  & Q(keV) & $E_P$(keV) & $J_{P}^{\pi}$ & $E_D$(keV) & $J_{D}^{\pi}$ & \multicolumn{1}{c}{Theo.}&\multicolumn{1}{c}{Expt.}&&\multicolumn{1}{c}{Theo.}&\multicolumn{1}{c}{Expt.} & Ref.\\
\hline
$^{42}$Sc   & $\beta^+$/EC &  6426.1  & 0.0   & $0^+$ &  0.0 &  $0^+$ & 3.493 & 3.485 & &  58.02  &99.98 & \cite{42sc}\\
$^{42}$Ti   & $\beta^+$/EC &  7016.48  & 0.0   & $0^+$ &  0.0 &  $0^+$ & 3.497 & 3.495 & &  19.73  & 47.74& \cite{42ti}\\
$^{46}$V   & $\beta^+$/EC & 7051.4   & 0.0   & $0^+$ &  0.0 &  $0^+$ & 3.497 & 3.484 & & 87.8   &99.97 & \cite{46v}\\
$^{50}$Mn  & $\beta^+$/EC & 7634.48   & 0.0   & $0^+$ &  0.0 &  $0^+$ & 3.497 & 3.485 & & 93.44   &99.93 & \cite{46v}\\
$^{48}$Fe   & $\beta^+$/EC & 11290   & 0.0   & $0^+$ &  3036.8 &  $0^+$ & 3.196 & 3.300 & &  26.94  & 35.03& \cite{48fe}\\
$^{50}$Fe   & $\beta^+$/EC & 8151   & 0.0   & $0^+$ &  0.0 &  $0^+$ & 3.497 & 3.490 & & 52.7   &74.17 & \cite{50fe}\\
$^{54}$Ni  & $\beta^+$/EC & 8790   & 0.0   & $0^+$ &  0.0 &  $0^+$ & 3.493 & 3.501 & & 51.13   &79.17 & \cite{54ni}\\
$^{58}$Zn  & $\beta^+$/EC & 9364   & 0.0   & $0^+$ &  203 &  $0^+$ & 3.497 & 3.486 & & 17.44   &72.08 & \cite{58zn}\\
\hline
\end{tabular}
\end{center}
\end{table*}
\begin{table*} 
\begin{center}
    \leavevmode
    \caption{ The theoretical (SM) results of $\beta$$^+$-decay half-lives for the concerned nuclei
    are compared with the experimental data, experimental $Q$ values, $I_{\beta^+}$ + $I_{\epsilon}$ - decay probabilities and theoretical quenched $\sum$ $B(GT)$ values are reported in this table.}

\begin{tabular}{lrccccccccc}
\hline
                   &              & $Q$ value      && Sum &\multicolumn{2}{c}{Half-life}     &$I_{\beta^+} + I_{\epsilon}$\\
\cline{5-6}
$^{A}$Z$_{i}(J^{\pi})$  & $^{A}$Z$_{f}$  & (keV) &$B(GT)$ &\multicolumn{1}{c}{Theo.}&\multicolumn{1}{c}{Expt.}&&
 $\%$\\
\cline{5-6}
\hline
$^{41}$Sc(${7/2}^-$)& $^{41}$Ca  & 6495.48$\pm$16   &2.436(2.110)    &   1213.8(1402.04) ms        &  596.3$\pm$17 ms\cite{41sc} &   &100\\
$^{42}$Sc($0^+$)    & $^{42}$Ca  & 6426.10$\pm$10   &0.485(0.420)    &   822.2(950.4) ms           &  680.79$\pm$28 ms\cite{42sc}  &       &100 \\
$^{43}$Sc($7/2^-$)  & $^{43}$Ca  & 2220.7$\pm$19    &0.064(0.055)    & 4.4(5.1) h                &   3.89$\pm$12 h\cite{43sc}   &             &100 \\
$^{42}$Ti($0^+$)    & $^{42}$Sc  & 7016.48$\pm$22   &2.528(2.188)    & 181.8(210.1) ms               &   208.65$\pm$80 ms\cite{42ti}   &             &100 \\
$^{43}$Ti($7/2^-$)  & $^{43}$Sc  & 6867$\pm$7       &1.892(1.638)    & 1054.7(1218.2) ms           &   509$\pm$5 ms\cite{43ti}   &             &100 \\
$^{45}$Ti($7/2^-$)  & $^{45}$Sc  & 2062.1$\pm$5     &0.517(0.448)    & 132.4(153.0) min            &   184.8$\pm$5 min\cite{45ti}   &             &100 \\
$^{44}$V($2^+$)     & $^{44}$Ti  & 1.343$\times 10^{4}$$\pm$12  & 0.804(0.696)   & 98(113.2) ms&   111$\pm$7 ms\cite{44v}   &             &100 \\
$^{45}$V($7/2^-$)   & $^{45}$Ti  &  7126$\pm$17     & 0.256(0.221)   & 2335(2697.1) ms&   547$\pm$6 ms\cite{45v}   &             &100 \\
$^{46}$V($0^+$)     & $^{46}$Ti  &  7051.4$\pm$10   & 0.183(0.159)   & 810.1(935.7) ms&   422.50$\pm$11 ms\cite{46v}   &             &100 \\
$^{47}$V($3/2^-$)   & $^{47}$Ti  &  2930.34$\pm$30  & 0.401(0.347)   & 23.9(27.6) min&   32.6$\pm$3 min\cite{47v}   &             &100 \\
$^{45}$Cr($7/2^-$)  & $^{45}$V  &  1.291$\times 10^4$$\pm$50  & 0.401(0.347)   & 110.4(127.5) ms&   60.9$\pm$4 ms\cite{45cr}   &             &100 \\
$^{47}$Cr($3/2^-$)  & $^{47}$V &   7444$\pm$14      &0.160(0.138)    & 2688.7(3105.6) ms&   500$\pm$15 ms\cite{47cr}   &             &100 \\
$^{48}$Mn($4^+$)    &$^{48}$Cr &  13525$\pm$10      &0.320(0.277)    &176.5(238.9) ms &   157.7$\pm$22 ms\cite{48mn}   &             &100 \\

$^{49}$Mn($5/2^-$)  &$^{49}$Cr  & 7715$\pm$24       &0.192(0.166)    &1848.1(2134.7) ms &   382$\pm$7 ms\cite{ENSDF}   &             &100 \\
$^{50}$Mn($0^+$)    &$^{50}$Cr &  7634.48$\pm$7     &0.041(0.035)    &592.5(684.3) ms &   283.19$\pm$10 ms\cite{46v}   &             &100 \\
$^{51}$Mn($5/2^-$)  &$^{51}$Cr  & 3207.5$\pm$3      &0.033(0.028)    &47.3(54.6) min &   46.2$\pm$1 min\cite{51mn}   &             &100 \\
$^{48}$Fe($0^+$)    &$^{48}$Mn  & 11290$\pm$90      & 1.397(1.209)   &11.1(12.7) ms  &   45.5$\pm$8 ms\cite{48fe}   &             &100 \\
$^{49}$Fe($7/2^-$)  &$^{49}$Mn  & 16895$\pm$73      & 0.204(0.177)   &267.8(309.4) ms &   64.7$\pm$3 ms\cite{49fe}   &             &100 \\
$^{50}$Fe($0^+$)    & $^{50}$Mn & 8151$\pm$8        & 1.220(1.056)   &241.7(279.2) ms &   152.0$\pm$6 ms\cite{50fe}   &             &100 \\
$^{51}$Fe($5/2^-$)  &$^{51}$Mn  & 8041$\pm$9        & 0.332(0.287)   &1221.1(1410.4) ms &   305$\pm$2 ms\cite{51fe}   &             &100 \\
$^{53}$Fe($7/2^-$)  &$^{53}$Mn &  3742.6$\pm$17     & 0.058(0.050)   &9.3(10.7) min &   8.51$\pm$2 min\cite{53fe}   &             &100 \\
$^{50}$Co($6^+$)    &$^{50}$Fe &  16895$\pm$73      & 0.051(0.040)   &255.3(294.4) ms &   38.8$\pm$2 ms\cite{50co}   &             &100 \\
$^{52}$Co($6^+$)    &$^{52}$Fe &  14340             & 0.354(0.306)   &163.1(188.4) ms &   104$\pm$7 ms\cite{52co}   &             &100 \\
$^{53}$Co($7/2^-$)  &$^{53}$Fe &  8300$\pm$18       & 0.298(0.258)   &977.6(1129.2) ms &   240$\pm$20 ms\cite{53co}   &             &100 \\
$^{55}$Co($7/2^-$)  &$^{55}$Fe &  3451.8$\pm$4      & 0.021(0.018)   & 9.9(11.5) h &   17.53$\pm$3 h\cite{ENSDF}   &             &100 \\
$^{57}$Co($7/2^-$)  &$^{57}$Fe &  836.0$\pm$4       & 0.002(0.001)  & 198.5(229.3) d &   271.74$\pm$6 d\cite{ENSDF}   &             &100 \\
$^{58}$Co($2^+$)    &$^{58}$Fe &  2307.6$\pm$12     &  0.002(0.001)  &44.4(51.3) d &   70.86$\pm$6 d\cite{ENSDF}   &             &100 \\
$^{54}$Ni($0^+$)    &$^{54}$Co  &  8.79$\times 10^3 $$\pm$5  &1.128(0.976)    &164.1(189.5) ms &   114.2$\pm$3 ms\cite{54ni}   &             &100 \\
$^{56}$Ni($0^+$)    &$^{56}$Co  &  2136$\pm$12  &0.192(0.165)    & 0.05(0.06) d &  6.07 $\pm$10 d\cite{ENSDF}   &             &100 \\
$^{57}$Ni($3/2^-$)  &$^{57}$Co &   3264.2$\pm$26    &0.029(0.025)    & 32.8(37.9) h &   35.60$\pm$6 h\cite{ENSDF}   &             &100 \\
$^{57}$Cu($3/2^-$)  &$^{57}$Ni &   8770$\pm$16      &1.656(1.769)    & 173.2(151.8) ms &   196.3$\pm$7 ms\cite{57cu}   &             &100 \\
$^{58}$Cu($1^+$)    &$^{58}$Ni &   8565.6$\pm$14    &1.216(1.298)    & 0.3(0.2) s &   3.204$\pm$7 s\cite{58cu}   &             &100 \\

$^{59}$Cu($3/2^-$)  &$^{59}$Ni &  4798.4$\pm$4   &0.271(0.289)    & 121.2(106.2) s &   81.5$\pm$5 s\cite{59cu}   &             &100 \\
$^{60}$Cu($2^+$)    &$^{60}$Ni &  6128.0$\pm$16  &0.095(0.101)    & 0.75(0.65) min &   23.7$\pm$4 min\cite{60cu}   &             &100 \\
$^{61}$Cu($3/2^-$)  &$^{61}$Ni &  2237.8$\pm$10  &0.194(0.207)    & 2.8(2.4) h &   3.336$\pm$10 h\cite{61cu}   &             &100 \\
$^{62}$Cu($1^+$)    &$^{62}$Ni &  3958.90$\pm$48 &0.094(0.101)    & 5.98(5.24) min &   9.67$\pm$3 min\cite{62cu}   &             &100 \\
$^{64}$Cu($1^+$)    &$^{64}$Ni &  1674.62$\pm$21 &0.051(0.055)    & 21.0(18.4) h &   12.7006$\pm$20 h\cite{64cu}   &             &100 \\
$^{58}$Zn($0^+$)    &$^{58}$Cu &  9364$\pm$50    &2.451(2.620)    &44.8(42.0) ms &   86$\pm$8 ms\cite{58zn}   &             &100 \\
$^{59}$Zn($3/2^-$)  &$^{59}$Cu &  9142.8$\pm$6   &1.425(1.522)    &137.3(120.3) ms &   178.6$\pm$18 ms\cite{59zn}   &             &100 \\
$^{60}$Zn($0^+$)    &$^{60}$Cu &  4170.8$\pm$16  &0.235(0.252)    & 4.2(3.7) min &   2.38$\pm$5 min\cite{60zn}   &             &100 \\
$^{61}$Zn($3/2^-$)  &$^{61}$Cu &  5635$\pm$16    &0.145(0.155)    &65.5(57.37) s &   89.1$\pm$2 s\cite{61zn}   &             &100 \\
$^{62}$Zn($0^+$)    &$^{62}$Cu &  1619.5$\pm$7   &0.152(0.163)    & 15.1(13.3) h &   9.193$\pm$15 h\cite{62zn}   &             &100 \\
$^{63}$Zn($3/2^-$)  &$^{63}$Cu &  3366.5$\pm$16  &0.093(0.100)    & 48.2(42.2) min &   38.47$\pm$5 min\cite{ENSDF}   &             &100 \\
$^{65}$Zn($5/2^-$)  &$^{65}$Cu &  1352.1$\pm$3   &0.012(0.013)    & 4.9(4.4) d &   243.93$\pm$9 d\cite{ENSDF}   &             &100 \\
\hline
\end{tabular}
\end{center}
\end{table*}


\section{Results and discussions}
 The computed Phase Space Factors (PSF) for $\beta^+$/EC decay along with log$(f^{(+)}_0+f^{(EC)}_{0})$ values are given in table 1. For higher $Q$ value, the relation $f^{(+)}_0$$\gg$$f^{(EC)}_0$ and the half-life of a $\beta^+$/EC transition is determined by the $\beta^+$ decay. For small decay energies the electron-capture dominates as reflected from the table. We have calculated  $f^{(+)}_0$ and $f^{(EC)}_0$ for small decay energies and compared them in table 1. We considered the calculated log$(f^{(+)}_0+f^{(EC)}_{0})$ values using Eq. 7 in our shell model calculations for 13  nuclei ($^{43}$Sc, $^{45}$Ti, $^{53}$Fe, $^{55}$Co, $^{57}$Co, $^{58}$Co, $^{56}$Ni, $^{57}$Ni, $^{61}$Cu, $^{64}$Cu, $^{60}$Zn, $^{62}$Zn and $^{65}$Zn) where $f^{(EC)}_0$ were significant. The shell model results of half-lives and branching ratios are improved for these nuclei after including the dominant EC phase space factor in our calculations.

The theoretical shell-model (SM) results of excitation energies as well as $\beta$-decay properties such as log$ft$ values, and branching percentages of the concerned nuclei are compared with the experimental data in table 2. 
 Columns 3 and 4 represent the theoretical and experimental excitation energies, respectively of each
state associated with the  $\beta^+$ - decays. Columns 5 and 6 represent quenched theoretical and experimental
log$ft$ values, respectively. The theoretical log$ft$ values are quenched by quenching factor q = 0.719 for $fp$ space and q = 0.743 for $f_{5/2}pg_{9/2}$ space, respectively. The theoretical and experimental branching fractions are listed in columns 7 and 8, respectively. The theoretical results of excitation energies, log$ft$ values, and the branching ratios are in reasonable agreement with the experimental data. 

Table 3 represents the comparison of theoretical and experimental $\beta$-decay half-lives of concerned nuclei. The experimental $\beta$-decay $Q$ values, 
$I_{\beta^+}$+$I_{\epsilon}$ decay probabilities and theoretical quenched $\sum B(GT)$ values are also reported.
The first and second columns represent parent and daughter nuclei, respectively. The experimental $Q$ values are taken from \cite{ENSDF} and listed in column 3. The theoretical quenched $\sum{B(GT)}$ values are presented in column 4. In column 4, the quenched $\sum{B(GT)}$ values are reported by quenching factor q = 0.719 and 0.743 for $fp$ and $f_{5/2}pg_{9/2}$ space, respectively, while in the small bracket, the theoretical results are quenched by quenching factor q = 0.669 and 0.768 for $fp$ and $f_{5/2}pg_{9/2}$ space, respectively. The quenched $\sum{B(GT)}$ values are smaller with q = 0.669 than 0.719 thus the quenched half-life with q = 0.669 are a little bit larger than q = 0.719 for $fp$ space. In the case of $f_{5/2}pg_{9/2}$ space, the quenched $\sum{B(GT)}$ values with quenching factor q = 0.768 are larger than q = 0.743 thus the quenched half-lives with q = 0.768 are a little bit smaller than q = 0.743 and are in general closer to the experiment.

 Columns 5 and 6 present theoretical and experimental $\beta$-decays half-lives, respectively. In column 5, the quenched half-lives are reported by quenching factor q = 0.719 and 0.743 for $fp$ and $f_{5/2}pg_{9/2}$ space, respectively, while in the small bracket, the theoretical results are quenched by quenching factor q = 0.669 and 0.768 for $fp$ and $f_{5/2}pg_{9/2}$ space, respectively. Overall the quenched half-life results with q = 0.719 are closer to the experimental data for $fp$ space while in case of $f_{5/2}pg_{9/2}$ space, the results with q = 0.768 are closer to the experimental data. The experimental $I_{\beta^+}$+$I_{\epsilon}$ decay probabilities are presented in the last column. In our present work the ground state spin parity $J^{\pi}$ of parent nuclei is determined using shell model for nuclei where the experiment ground state $J^{\pi}$ is not confirmed, it is indicated by * in table 2.

For $^{41}$Sc, $^{43}$Sc, $^{42}$Ti nuclei experimental data are available in refs. \cite{41sc,43sc,42ti} and the experimental $\beta^+$-decay half-lives for these nuclei are found to be 596.3 $\pm$ 17 ms, 3.891 $\pm$ 12 h, and 208.65 $\pm$ 80 ms, respectively, whereas the SM results with q = 0.719 are 1213.8 ms, 4.4 h and 181.8 ms, respectively. The quenched SM results are closer to the observed values. In the case of $^{43}$Sc, the $f^{(EC)}_{0}$ values for the transitions at $7/2^-$ (0 MeV) and $5/2^-$ (0.373 MeV) are significant compared to $f^{(+)}_{0}$. The quenched half life is 4.4 h with q = 0.719 using computed log$(f^{(+)}_0+f^{(EC)}_{0})$ values, where as without $f^{(EC)}_{0}$ the half-life is 4.6 h, also the branching ratios shifted towards experimental value after including $f^{(EC)}_{0}$ in calculation.

The half-life of $^{43}$Ti was found to be 509$\pm$5 ms in ref. \cite{43ti} which is determined from the high-energy $\gamma$-rays of mass-separated samples, while the calculated shell model result is 1054.7 ms with q = 0.719.
 For $^{45}$Ti, the $f^{(EC)}_{0}$ is significant at the transitions $7/2^-$ (0 MeV), $5/2^-$ (0.543 MeV), $5/2^-$ (0.720 MeV) and $7/2^-$ (1.408 MeV) with log$(f^{(+)}_0+f^{(EC)}_{0})$ values 0.652, -0.287, -0.614, and -0.932, respectively. The predicted half-life is 132.4 min with quenching value 0.719. The branching ratios are also improved.

The $^{48,49}$Fe and $^{50}$Co nuclei are proton-rich nuclei \cite{44v} which are produced by fragmentation of a $^{58}$Ni beam at 650 MeV/u with the GSI Projectile-Fragment Separator FRS. The theoretical quenched $\beta^+$-decay half-lives, quenched log$ft$ values and branching ratios of these nuclei are also calculated and compared with the experimental data.
 There are five transitions in $^{53}$Fe at $7/2^-$ (0 MeV), 
$5/2^-$ (0.378 MeV), $9/2^-$ (1.619 MeV), and $5/2^-$ (2.273 MeV) where the EC decay is significant. The SM result of half-life is 9.3 min with q = 0.719 which is close to the experimental value 8.51 min.

In the case of $^{42}$Sc $\rightarrow$ $^{42}$Ca, $^{42}$Ti $\rightarrow$ $^{42}$Sc, $^{46}$V $\rightarrow$ $^{46}$Ti, $^{50}$Mn $\rightarrow$ $^{50}$Cr, $^{48}$Fe $\rightarrow$ $^{48}$Mn, $^{50}$Fe $\rightarrow$ $^{50}$Mn, $^{54}$Ni $\rightarrow$ $^{54}$Co, and $^{58}$Zn $\rightarrow$ $^{58}$Cu transitions, the Fermi matrix elements are non-zero. So, we reported the superallowed Fermi decay for $0^+$ $\rightarrow$  $0^+$ transition for these nuclei and compared the calculated log$ft$ values and branching ratios with experimental data in table 3. The calculated half-lives and branching ratios of these nuclei improved after including the Fermi transition. There are notable differences observed between calculated and experimental log$ft$ values for $^{53}$Fe,  $^{55}$Co and $^{60}$Zn decay, probably the results may be improved if we take extended model space such as $fpg_{9/2}$.
 There are six transitions observed in $^{55}$Co at $5/2^-$ (0.931 MeV), $7/2^-$ (1.316 MeV), $7/2^-$ (1.408 MeV), $5/2^-$ (2.144 MeV), $9/2^-$ (2.212 MeV), and $9/2^-$ (2.301 MeV) in which the EC branching ratios are significant. Our calculations in table 1 also support that the $f^{(EC)}_{0}$ is significant for these transitions. After using the calculated log$(f^{(+)}_0+f^{(EC)}_{0})$ values the SM result of half-life is 9.9 h with q = 0.719, where as without the $f^{(EC)}_{0}$, the SM result is 1.36 h.
In case of $^{57}$Co, 99.8$\%$ EC transition is observed at $5/2^-_1$ (0.136 MeV) with half-life 271.7 d. The calculated $f^{(EC)}_{0}$ $>$ $f^{(+)}_{0}$ and the calculated SM results with quenching factor 0.719 is 275.1 d which is close to the experimental value 271.74 d.
In case of $^{58}$Co the dominant EC transitions are observed, our calculations also show that $f^{(EC)}_{0}$ is dominating over  $f^{(+)}_{0}$, using log$(f^{(+)}_0+f^{(EC)}_{0})$ from table 1 the SM half-life is 45.0 d with q = 0.719. 

In $^{56}$Ni decay, the 100$\%$ EC transition was observed at $1^+$ (1.720 MeV) with log$ft$ = 4.40. When the $Q$ values of the EC transition are less than the electron rest mass energy the endpoint energy $E_0$ becomes negative and the $\beta^+$ mode cannot exist. As in table 1, the $f^{(+)}_{0}$ is negative for $^{56}$Ni decay (indicated by ``--") with $f^{(EC)}_{0}$ = 0.034. The calculated half-life is 6.9 d by considering only EC transition which is very close to the experimental value 6.07 d.

 In case of $^{61}$Cu we computed and compared PSF for EC and $\beta^+$-decay and reported the half-life 2.8 h with q = 0.719 which is close to the experimental value 3.336 h. Similarly we computed $f^{(EC)}_{0}$ and $f^{(+)}_{0}$ for $^{64}$Cu and reported half-life 21.0 h with q = 0.719.
In case of $^{64}$Cu decay the observed $\beta^+$/EC -decay branching ratio 61.5 $\%$ is scaled to 100 $\%$ in order to be consistently comparison between computed $\beta^+$/EC -decay branching.
 The SM half-life results slightly change for $^{60}$Zn, $^{62}$Zn and $^{65}$Zn after considering $f^{(EC)}_{0}$ in SM calculations.

The theoretical and experimental $\beta$-decay half-lives of concerned nuclei are plotted in Figs. 2 - 3 and the data are taken from table 4. In the figure, we used a log frame to show the $\beta^+$-decay half-lives. The figure indicates that the $\beta$-decay half-lives increase rapidly with the increasing mass number. The experimental data are presented by dotted lines with error bars. For most of the $fp$ and $f_{5/2}pg_{9/2}$ shell nuclei SM results of $\beta$-decay half-lives are in reasonable agreement with the experimental data.
The SM results of  Cu and Zn isotopes for $f_{5/2}pg_{9/2}$ space using JUN45 interaction, show a reasonable agreement with experimental data. 

\section{Summary and Conclusion}
In the present work, we have reported a comprehensive nuclear shell model study of $\beta$-decay half-lives, log$ft$ values, and branching fractions
for the $fp$ and $f_{5/2}pg_{9/2}$ shell nuclei with $Z = 21 -30$. The calculations have been performed in two different model spaces. For $fp$ shell nuclei the KB3G effective interaction has been used and for Cu and Zn nuclei in $f_{5/2}pg_{9/2}$ model space the JUN45 effective interaction has been used. Over all the calculated results of excitation energies, log$ft$  values, half-lives, and branching fractions for most of the nuclei are in
reasonable agreement with the available experimental data.
The present shell model results of $\beta$-decay, half-lives, log$ft$ values, and branching fractions will add more information to the earlier experimental works.
Further, the calculated quenching factors in this work for $fp$ and $f_{5/2}pg_{9/2}$ space will play an important role for $\beta$-decay study in this space.
\section*{Acknowledgment:}
V. Kumar acknowledges financial support from SERB Project (File No. EEQ/2019/000084), Govt. of India.
The authors also acknowledge PARAM Shivay Computing facility at IIT (BHU) Varanasi. V. Kumar also acknowledges financial support from IoE Seed Grant, BHU (R/Dev/D/IoE/Seed Grant-II/2021-22/39960).
PCS acknowledges financial support from SERB (India), CRG/2022/005167. 
In addition, we would like to thank the National Supercomputing Mission (NSM) for providing computing resources of ‘PARAM Ganga’ at the Indian Institute of Technology Roorkee, implemented by C-DAC and supported by the Ministry of Electronics and Information Technology (MeitY) and Department of Science and Technology (DST), Government of India. We would like to thank Prof. T. Suzuki for useful suggestions.

\input{bibliography}

\end{document}

%% file: bibliography.tex
\bibliography{utphys}
\bibliography{references}